%% file: Main.tex
\documentclass{article}
\usepackage{spconf,amsmath,graphicx,dirtytalk,booktabs,xcolor}
\usepackage{graphicx}
\usepackage{pgfplots}
\usepackage{caption,subcaption}
\usepackage{dblfloatfix}

\def\e{\mathbf{e}}
\def\x{\mathbf{x}}
\def\y{\mathbf{y}}
\def\yi{\mathbf{y}}
\def\yifuzz{\mathbf{y}^{\text{fuzz}}}
\def\yl{\mathbf{y}^l}
\def\ystar{\mathbf{y}^*}
\def\yoner{\mathbf{y}_1^r}
\def\yNr{\mathbf{y}_N^r}

\usepackage{moredefs} 
\usepackage{color}

\DeclareMathOperator*{\argmax}{argmax}

\title{Improving Proper Noun Recognition in End-To-End ASR by Customization of the MWER Loss Criterion}
\name{Cal Peyser, Tara N. Sainath, Golan Pundak}
\address{Google Inc. \\
\fontsize{9}{9}\selectfont\ttfamily\upshape
\{cpeyser,tsainath,golan\}@google.com}

\begin{document}
\ninept
\maketitle

\input{abstract}
\input{intro}

\input{background}
\input{mwer}
\input{experiments}
\input{results}
\input{conclusions}

\section{Acknowledgements}
The authors would like to thank Ke Hu and Rohit Prabhavalkar for helpful comments regarding this work.

\vfill\pagebreak
\bibliographystyle{IEEEbib}
\bibliography{Main}

\end{document}

%% file: abstract.tex
\begin{abstract}
	Proper nouns present a challenge for end-to-end (E2E) automatic speech recognition (ASR) systems in that a particular name may appear only rarely during training, and may have a pronunciation similar to that of a more common word.  Unlike conventional ASR models, E2E systems lack an explicit pronounciation model that can be specifically trained with proper noun pronounciations and a language model that can be trained on a large text-only corpus. Past work has addressed this issue by incorporating additional training data or additional models.  In this paper, we instead build on recent advances in minimum word error rate (MWER) training to develop two new loss criteria that specifically emphasize proper noun recognition.  Unlike past work on this problem, this method requires no new data during training or external models during inference.  We see improvements ranging from 2\% to 7\% relative on several relevant benchmarks.
\end{abstract}
%

%% file: intro.tex
%
\section{Introduction}
\label{sec:intro}

A common challenge in creating a user-friendly speech recognition experience is to improve performance on \say{tail} utterances.  Specifically, certain classes of speech are inherently more difficult for speech systems by virtue of their ambiguity, rareness in training,  or unusual verbalization.  Examples include accented speech \cite{Vu14, Shor19, Jain18}, cross-lingual speech \cite{Schultz00, Hu19}, and numerics \cite{Vasserman15, Peyser19}.

Proper nouns form such a category of tail utterances that is both important for ASR quality and difficult to get right.  Consider, for example, the hypothetical voice search query \say{Directions to Beaumont}.  The proper noun \say{Beaumont} could be easily confused with the similar-sounding and semantically plausible alternatives \say{Hallmark} or \say{Walmart}, entirely changing the meaning of the utterance.  Also, since \say{Beaumont}, which is the name of a city in Texas, might occur only rarely in a typical training corpus, an ASR model may consider it less likely than the names of the national retail chains \say{Walmart} and \say{Hallmark}.

In conventional ASR systems, which are split into an acoustic model (AM), a pronunciation model (PM), and a language model (LM), optimizing the PM directly has been shown to improve tail performance by injecting knowledge of the pronunciation of proper nouns.  The authors of \cite{Beaufays03} augment the PM with pronunciations of names, achieving improved performance on a test set derived from a corporate directory of names.  Similarly, an approach is proposed in \cite{Laurent10} in which the pronunciations of proper nouns in foreign languages are mined using a machine translation system.  Conventional models also have the advantage of a LM trained on a very large text-only corpus which can have greater exposure to proper nouns than a system trained only on audio-text pairs.

In recent years, end-to-end (E2E) ASR systems such as RNN-T \cite{Graves12} and LAS \cite{Chan15} have proven to be viable alternatives to conventional speech systems, especially in the on-device setting where memory is limited \cite{He18}.  These systems achieve strong results with small model size by folding the AM, PM and LM into a single neural model.  While this simplifies the model architecture, it removes the explicit PM as a site for the injection of knowledge into the model, making it more difficult to model specific requirements like proper noun pronunciation.  It also reduces the exposure of the model to proper nouns, since unlike the LM of a conventional model an E2E system is trained entirely on audio-text pairs.

Several directions have been explored for improving proper noun recognition in E2E models.  In \cite{Alon19}, phonetic fuzzing of proper nouns in training an E2E system yields improvements on several proper noun benchmarks, but the approach is restricted to the biasing case, where candidate proper nouns are passed directly to the model.  In addition, several techniques incorporating text-only data into an E2E model, including shallow fusion, cold fusion, and deep fusion have been proposed and have been shown to improve performance on tail utterances \cite{Toshniwal18, Kannan18}.  However, this approach requires an external language model to be used in both training and inference.

In an attempt to make up for the lack of an explicit LM potentially trained on billions of words, there has also been work on including large corpora of unsupervised audio or text data during training of an E2E model.  The authors of \cite{Li19} use a conventional model to decode unsupervised utterances that contain proper nouns and train an E2E system on the resulting transcripts.  Conversely, the authors of \cite{He19} apply a text-to-speech (TTS) system to unsupervised text data and train an E2E system on the resulting audio.  In \cite{Karita19} and \cite{Hori19}, unsupervised data is included in an E2E model though a seperately trained model with a reconstruction loss.  While all of these approaches yield improvements, they require use of external datasets.

In this paper, we explore approaches to proper noun recognition in E2E systems that require neither additional data in training nor additional external models during inference by using loss functions that specifically target proper nouns during training.  The problem of task-specific losses is explored in \cite{Prabhavalkar17} and \cite{Weng18}, in which minimum word error rate (MWER) training \cite{Shannon17} is adapted to E2E models to optimize word error rate (WER) directly instead of a differentiable surrogate like cross-entropy loss.  This is done by computing WER scores on the results of a beam search.  In \cite{Sainath19}, a second-pass LAS model is optimized by MWER sampling from a beam search of the first-pass RNN-T model, demonstrating that MWER training can be used to transfer knowledge from one model to another.  We seek to further specialize MWER training to improve performance on tail utterances.

We propose two new methods.  In the first, we use an entity-tagging system as in \cite{Huang15} to identify proper nouns in ground truth transcripts and increase the loss for hypotheses that miss a proper noun during MWER training.  In the second, we draw inspiration from \cite{Alon19} and \cite{Sainath19} and add additional hypothesis to the MWER beam in which proper nouns have been replaced by phonetically similar alternatives.  This provides the model with knowledge of those possible mistakes and alternative spellings during training.  We show that these methods improve performance over unmodified MWER training on proper noun test sets by as much as 7.2\% relative.

The rest of this paper is organized as follows.  Section \ref{sec:background} provides background on MWER training of a two-pass E2E ASR system.  Section \ref{sec:mwer-customizations} describes our two new MWER training strategies for improving proper noun recognition.  Section \ref{sec:experiments} outlines our experimental setup and Section \ref{sec:results} presents results.  Finally, Section \ref{sec:conclusions} concludes by summarizing our work.

%% file: background.tex
\section{Background}
\label{sec:background}

In this section, we summarize the two-pass E2E ASR system used in this work, as well as the MWER scheme used to fine-tune it during training.  See \cite{Sainath19} for further details on the model, training procedure, and MWER fine-tuning method.

\begin{figure}[htb]
\centering
  \includegraphics[scale=0.4]{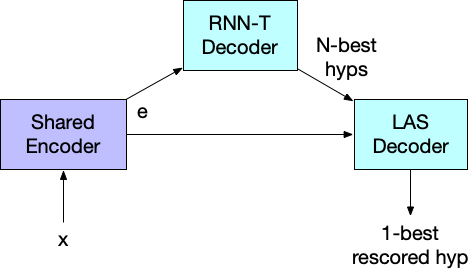}
  \caption{Two-Pass E2E Architecture, adapted from \cite{Sainath19}.}
  \label{fig:TwoPass}
  \vspace{-0.2in}
\end{figure}

\subsection{Two-Pass E2E Model}
\label{ssec:model-arch}
We will first describe the architecture of the two-pass system (Figure \ref{fig:TwoPass}) and how it is used during inference.  First, a single encoder maps acoustic features $\x = (x_1,...,x_T)$ to frame-by-frame embeddings $\e = (e_1,...,e_T)$, where each $e_i$ is a 512-dimensional vector.  These embeddings are consumed via an attention mechanism by an RNN-T decoder that produces word piece output.  A beam search is run with the RNN-T decoder to produce an N-best list $B_{\text{RNN-T}} = (\yoner, ..., \yNr)$ of hypotheses.

A second, LAS decoder attends to the encoder output and computes:
\begin{equation}
\yl = \argmax_{\y \in B_{\text{RNN-T}}} P(\y | \e)
\end{equation}

That is, the LAS decoder acts as a rescorer, choosing the single RNN-T hypothesis with maximum likelihood from the N-best list.  In \cite{Sainath19}, a second mode is explored in which the LAS decoder emits word piece output, and where the final $\yl$ is determined by beam search. It was found in \cite{Sainath19} that rescoring gives better inference speed than beam search, so we use a rescoring model in this work.

\subsection{Training Procedure}
\label{ssec:training-procedure}
Model training consists of three steps.  First, the shared encoder and RNN-T decoder are trained as in \cite{Graves12}.  Second, the LAS decoder is trained without updating the parameters of the shared encoder or the RNN-T decoder.  During this step, the LAS decoder is trained with a cross-entropy, teacher-forcing loss.

The final training step is MWER training as described in \cite{Prabhavalkar17}.  This procedure is designed to focus training on the WER objective by using as a loss the weighted average of word errors in an N-best beam of hypotheses.  Applying that idea directly to our architecture, we might fine-tune our second, LAS decoder with the following loss:
\begin{equation}
L(\x, \ystar) = \sum_{\y \in B_{\text{LAS}}} P(\y | \x) \hat{W}(\yi, \ystar)
\end{equation}
\\
where $\ystar$ is the ground truth, $B_{\text{LAS}}$ is an N-best list of hypothesis drawn from the LAS decoder using a beam search, $P(\y | \x)$ is the normalized posterior for the hypothesis $\y$, and $\hat{W}(\y, \ystar)$ gives the difference between the number of word errors in the hypothesis $\y$ and the average number of word errors across the beam. However, since the LAS decoder will be used during inference as a rescorer, we seek instead to optimize its ability to assign high likelihood to the best RNN-T hypotheses.  Therefore, we modify the loss:
\begin{equation}
L(\x, \ystar) = \sum_{\y \in B_{\text{RNN-T}}} P(\y | \x) \hat{W}(\yi, \ystar)
\end{equation}
\\
where $B_{\text{RNN-T}}$ is obtained by beam search on the RNN-T decoder as defined above.  The operative insight in this phrasing of the MWER loss criterion is that the set of hypotheses used does not need to arise directly from the model being optimized.  Rather, it simply must represent a distribution over which the model should learn to assign probability mass.  It is this idea that we build on with beam modification in Section \ref{ssec:beam-modification} below.

%% file: mwer.tex
\section{MWER Customizations}
\label{sec:mwer-customizations}
In this section we present two modifications to the MWER fine-tuning loss which aim to improve recognition of proper nouns.

\subsection{Proper Noun Loss Augmentation}
\label{sec:proper-noun-loss-augmentation}
We define a new MWER fine-tuning loss:
\begin{equation}
L_{\text{Aug}}(\x, \ystar) = \sum_{\yi \in B_{\text{RNN-T}}} P(\yi | \x)\hat{W}(\yi, \ystar) \cdot C_\lambda(\yi, \ystar)
\end{equation}
\\
where 

\[
    C_\lambda(\yi, \ystar)= 
\begin{cases}
    \lambda,& \text{if $\ystar$ contains a proper noun that is not in $\yi$}\\
    1,              & \text{otherwise}
\end{cases}
\]
\\
for some constant $\lambda > 1$.  Proper nouns in each ground-truth transcription are tagged ahead of time by a separate system. We define a hypothesis $\y$ as containing a word sequence $P$ if the entire sequence $P$ occurs in order in $\y$.  For example, the proper noun \say{Cedar Rapids} is contained in the hypothesis \say{Population of Cedar Rapids}, but not in the hypothesis \say{Cedar tree height} or \say{Cedar Rapidss}.

This loss is designed to emphasize proper noun performance in training by increasing the penalty applied to the model when it assigns high probability to a hypothesis that misses a proper noun correctly.  A potential drawback of this approach is that increasing the gradient originating from proper noun errors will dilute the contribution of other error types.  For this reason, the hyperparameter $\lambda$ must be chosen carefully so as to balance proper noun recognition with performance on general utterances.

\subsection{Beam Modification with Fuzzing}
\label{ssec:beam-modification}
In Section \ref{ssec:training-procedure}, we saw how MWER training allows us to train a model to distribute probability density correctly between hypotheses that were not necessarily derived from the model itself.  We use this flexibility to optimize our model's ability to distinguish between proper nouns and phonetically similar, incorrect alternatives by adding those possible mistakes to the beam.  If the model assigns high likelihood to these mistakes, it is penalized accordingly.  We generate these alternatives by phonetic fuzzing, which gives the advantage of possibly introducing new words and spellings that may not have been emphasized in the training data.

For hypothesis $\yi \in B_{\text{RNN-T}}$ and corresponding ground truth $\ystar$, we define the operation FUZZ such that:

\[
	\text{FUZZ}(\yi, \ystar) =
\begin{cases}
	\yifuzz,& \text{if $\ystar$ and $\yi$ share a proper noun $P$}\\
	\yi,& \text{otherwise}
\end{cases}
\]
\\
We construct $\yifuzz$ by copying $\yi$, and then replacing the occurrence of the proper noun $P$ with a phonetically similar alternative, called a \say{fuzz}.  We then define the loss by combining the original beam from RNN-T with a fuzzed copy:
\begin{equation} \label{eq:fuzz}
L_{\text{Fuzzed}}(\x, \ystar) = \sum_{\yi \in B_{\text{RNN-T}} \: \cup \: \text{FUZZ}(B_{\text{RNN-T}}) } P(\yi | \x)\hat{W}(\yi, \ystar)
\end{equation}
\\
where $\text{FUZZ}(B_{\text{RNN-T}})$ is a beam composed of $\text{FUZZ}(\yi, \ystar)$ for each $\yi \in B_{\text{RNN-T}}$ and where $P(\yi | \x)$ gives the renormalized posterior that accounts for the doubling of the beam size.  We also define a hyperparameter $0 \le \tau \le 1$ that gives the probability that we use the loss $L_\text{Fuzzed}$.  Otherwise, we use the standard loss $L$ as defined in Section \ref{ssec:training-procedure}.  Twenty five fuzzes are computed ahead of time for each proper noun occurring in the training data.  This number is chosen to ensure diversity of fuzzes while limiting computational expense.  Each time a hypothesis $\yifuzz$ is generated during training, one of these fuzzes is chosen at random.

%% file: experiments.tex
\section{Experiment Setup}
\label{sec:experiments}

This section details our experimental setup.

\subsection{Data Preparation}
\label{ssec:data-preparation}
Our training set consists of anonymized, hand-transcribed utterances representing Google voice search traffic as in \cite{Arun19}.  We create multi-style training (MTR) data by adding noise derived from YouTube videos and from daily life environmental recordings. This noise is added at an average SNR of 12dB, according to the procedure in \cite{Kim17}.

We choose four test sets for this experiment, shown in Table \ref{tab:eval-sets}: one standard voice search benchmark, and three test sets specifically measuring proper noun performance. For the proper noun sets, common names, top songs, and popular app names are mined from the internet and templated into transcripts (e.g. \say{Play \$SONG}), after which audio is synthesized using an HMM-based TTS system like \cite{Gonzalvo16}.  More details on these test sets can be found in \cite{Pundak18}.  

{ 
	\begin{table}
	\centering
	\setcounter{table}{0}		
	\begin{tabular}{|c|c|c|c|}
	\toprule
	Test Set & \# Utts & OOV Rate & Rare Word Rate \\
	\midrule
	VOICE SEARCH & 14877 & 0.154 & 0.250 \\
	\midrule
	CONTACTS & 15416 & 0.048 & 0.252 \\
	\midrule
	SONGS & 15016 & 0.020 & 0.209 \\
	\midrule
	APPS & 16262 & 0.026 & 0.211 \\
	\bottomrule
	\end{tabular}
	\caption{The eval sets used in this work. Rare Word Rate refers to the proportion of words occurring less than 1,000 times in the training vocabulary.}
	\label{tab:eval-sets}
        \vspace{-0.2in}
	\end{table}
}

\subsection{E2E Model}
\label{ssec:e2e-model}
Our model is a two-pass model as described above and in \cite{Sainath19}.  The encoder is an 8-layer LSTM stack, each layer of which contains 2048 nodes, followed by 640 projection units.  The first decoder is an RNN-T decoder as in \cite{Graves12}, with a projection network and joint network.  The projection network is a 2-layer LSTM stack, also with 2048 nodes per layer and 640 projection units, while the joint network has 640 hidden units and projects into a word piece vocabulary of size 4096.  The second decoder is a LAS decoder as in \cite{Chan15} with an 2-layer LSTM stack with 2048 nodes per layer, projecting into the same word piece vocabulary.  In total, the model contains about 175 million parameters.  Encoder input is 128-dimensional log-Mel filterbank features using a 32ms window with a shift of 10ms.  We use the procedure in \cite{Pundak16} of stacking two frames before any given frame.  The model is implemented in Tensorflow and is trained in parallel on 16 TPUs.

\input{diff_analysis}

\begin{table*}[h!]{}
\begin{minipage}{0.65\textwidth}
	\setcounter{table}{1}	
	\begin{tabular}{|c|c|c|c|c|c|c|c|c|}
	\toprule
	& Baseline & $\lambda = 1.5$& $\lambda=3$ & $\lambda=4.5$ & $\lambda=10$ & $\lambda=100$ \\
	\midrule
	VS & 5.6& 5.6& \textbf{5.5} & 5.6 & 5.7 & 6.1 \\
	\midrule
	CONTACTS & 22.7 & 22.2 & \textbf{22.1} & 22.2 & 22.3 & 23.0 \\
	SONGS & 11& 10.7 &  \textbf{10.7} & 10.7 & 10.8 & 11.8 \\
	APPS & 7.5 & 7.2 &  \textbf{7.1} & 7.3 & 7.4 & 8.1 \\
	\bottomrule
	\end{tabular}
	\caption{Proper Noun Loss Augmentation WER (\%)}
	\label{tab:aug-results}
        \vspace{-0.2in}
\end{minipage}
\begin{minipage}{0.31\textwidth}
	\setcounter{table}{2}	
	\begin{tabular}{|c|c|c|c|}
	\toprule
	$\tau = .1$ & $\tau=.25$ & $\tau = .5$ & $\tau =1$  \\
	\midrule
	 5.9 & 5.9 & 5.8 & \textbf{5.8} \\
	\midrule
	22.4 & 22.5 & 22.4 & \textbf{22.3} \\
	10.3 & 10.3 & 10.3 & \textbf{10.2} \\
	\textbf{7.3} & 7.5 & 7.4 & 7.4 \\
	\bottomrule
	\end{tabular}
	\caption{Fuzzing WER (\%)}
	\label{tab:fuzzing-results}
        \vspace{-0.2in}
\end{minipage}
\end{table*}

\subsection{Proper Noun Tagging and Fuzzing}
\label{ssec:tagging-and-fuzzing}
We adopt a procedure similar to \cite{Alon19} for proper noun tagging and fuzzing.  We preprocess our dataset in two steps.  First, the transcript of each utterance is passed through a part-of-speech (POS) tagger to identify proper nouns.  The POS tagger is provided by the Google Cloud Natural Language API which implements an RNN model as in \cite{Huang15}.

We then perform a lookup in a prepared dictionary mapping n-grams to phonetic fuzzes.   In preparing this dictionary, we seek for a given n-gram likely mistakes that an ASR model would make.  To that end, we use a conventional ASR system to decode a large set of utterances, and track all n-grams occurring in hypotheses for the same utterance.  For an n-gram $n$, phrases that are often hypothesized by the model for the same utterance as $n$ are considered likely mistakes.  We then filter this list by phonetic similarity as in \cite{Hixon11} to obtain the final list of fuzzes.

%% file: diff_analysis.tex
\begin{figure*}[h!]
\centering
\begin{subfigure}[b]{0.4 \textwidth}
	\begin{tikzpicture}[scale=0.8]
	\begin{axis}[
	axis lines=middle,
	ymin=-0.02,
	ymax=0.0045,
	xmin=0,
	x label style={at={(axis description cs:0.6,-0.03)}},
        y label style={at={(axis description cs:-0.13,.5)},rotate=90,anchor=south},	
	title=,
	  legend pos=south east,
	  xlabel=Vocabulary,
	  ylabel=VWER Difference,
	  xticklabel style = {rotate=30,anchor=east},
	   enlargelimits = false,
	  xticklabels from table={vs.dat}{Occurrences},xtick=data]
	\addplot[orange,thick,mark=square*] table [y=AugDiff,x=X]{vs.dat};
	\addlegendentry{Loss Aug}	
	\addplot[green,thick,mark=triangle*] table [y=ModDiff,x=X]{vs.dat};
	\addlegendentry{Fuzzing}	
	\end{axis}
	\end{tikzpicture}
	\caption{VOICE SEARCH}
\end{subfigure}
\begin{subfigure}[b]{0.4\textwidth}
	\begin{tikzpicture}[scale=0.8]
	\begin{axis}[
	axis lines=middle,
	ymin=-0.0035,
	ymax=0.0071,
	xmin=0,
	x label style={at={(axis description cs:0.6,-0.03)}},
        y label style={at={(axis description cs:-0.1,.5)},rotate=90,anchor=south},		
	title=,
	  legend pos=south east,
	  xlabel=Vocabulary,
	  ylabel=VWER Difference,
	  xticklabel style = {rotate=30,anchor=east},
	   enlargelimits = false,
	  xticklabels from table={songs.dat}{Occurrences},xtick=data]
	\addplot[orange,thick,mark=square*] table [y=AugDiff,x=X]{songs.dat};
	\addplot[green,thick,mark=triangle*] table [y=ModDiff,x=X]{songs.dat};
	\end{axis}
	\end{tikzpicture}
	\caption{SONGS}
\end{subfigure}
	\caption{Plot of the difference between baseline and model vocabulary word error rate for subsets of the training vocabulary ranging from the entire vocabulary to words occurring at least 100,000 times for the VOICE SEARCH and SONGS test sets.}
\label{fig:diff-fig}
\end{figure*}
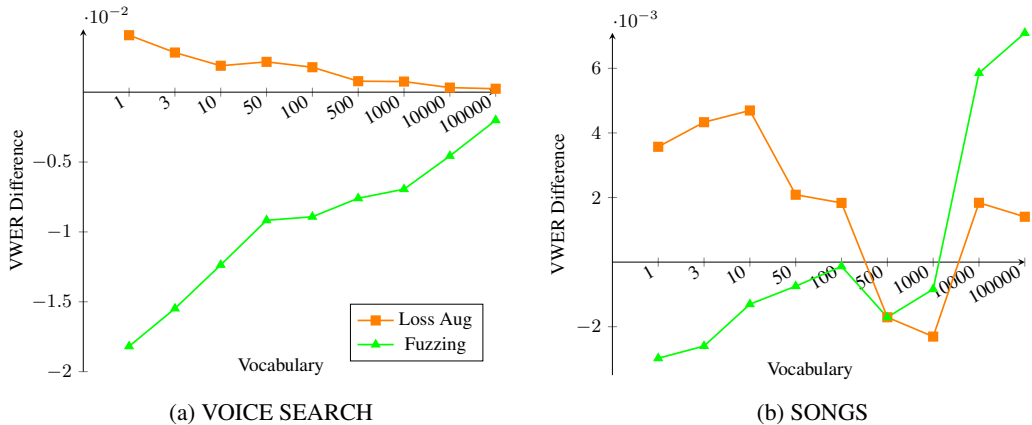

%% file: results.tex
\section{Results}
\label{sec:results}
In this section, we present the results of our experiments on several benchmarks, and provide analysis of error patterns.

\subsection{Loss Augmentation}
\label{ssec:loss-augmentation-results}
We ran several experiments, varying the hyperparameter $\lambda$ as described in Section \ref{sec:proper-noun-loss-augmentation}.  The results are presented in Table \ref{tab:aug-results}.  We see immediate improvement on the proper noun benchmarks with $\lambda= 1.5$ and achieve our best results for $\lambda = 3$, for which we see gains ranging from 2\% relative on VOICE SEARCH to 5\% relative on APPS compared to the baseline model.  Increasing $\lambda$ further reduces gains, and very large values of $\lambda$ in fact lead to deterioration on all test sets.  This is unsurprising, as concentrating too much loss on proper nouns is likely to dilute the loss assigned to other errors. 

\subsection{Beam Modification}
\label{ssec:beam-modification-results}
Results are presented in Table \ref{tab:fuzzing-results} for varying values of the hyperparameter $\tau$.  We generally see the best results for $\tau = 1.0$, that is, when $L_\text{Fuzzed}$ is used in every batch.  While we see the best results yet on the SONGS test set (7\% relative improvement over the baseline), we do not see the same improvement for the other benchmarks as we see with loss augmentation, and we in fact see deterioration on our VOICE SEARCH benchmark. We suggest a reason for this deterioriation in the next section.

Note that in this study, we do not combine loss augmentation and beam fuzzing in any experiment.

\subsection{Analysis}
\label{ssec:results-analysis}
Since our methods emphasize proper nouns, which typically are infrequently seen in training, we expect them to improve performance specifically on rare words.  To test this hypothesis, we define the \say{vocabulary word error rate} (VWER) of a model for a given vocabulary and test set as number substitution errors on words in the vocabulary divided by the total number of occurrences of vocabulary words in the test set. We consider specifically the difference between the VWER of the baseline and our models with subsets of the training vocabulary ranging from the entire vocabulary to the subset of words that occur at least 100,000 times.  As we increase this threshold of occurrences, we measure the performance difference on less and less rare words.  When this difference is positive, it measures the degree to which improvement on the subset of the vocabulary in question explain WER gains.  When it is negative, it similarly measures the source of degradation.  In Figure \ref{fig:diff-fig}, we plot this difference for the VOICE SEARCH and SONGS benchmark.

We find our expectation to be correct for the loss augmentation model.  As we restrict the vocabulary to more common words, the difference in performance between the baseline model and the augmentation model shrinks, suggesting that the improvement we see with loss augmentation is mostly attributable to better performance on rare words.  However, we are surprised to see the opposite trend in the fuzzing model.  The degradation we observe on the VOICE SEARCH benchmark is worse on rare words than on common words, and the large gains we see on the SONGS set seem to arise entirely from improvements on very common words. 

One possible explanation for these results is that $L_{\text{FUZZ}}$ as defined in equation \ref{eq:fuzz} doubles the beam by copying hypothesis that do not share a proper noun with the reference.  While this is done because of implementation constraints (TPU operations require a fixed input size), it may distort the $P(\yi|\x)$ term in $L_{\text{FUZZ}}$, since after normalization the copied hypothesis will have twice the probability mass than it would in the unmodified MWER criterion.  Since many rare words are proper nouns which would be fuzzed instead of copied, it's possible that this causes rare words to be de-emphasized during training.   Future research will assess fuzzing losses that don't copy hypothesis in order to realize the benefits of the approach on rare as well as common words.

%% file: conclusions.tex
\vspace{-0.1in}
\section{Conclusions}
\label{sec:conclusions}
In this paper, we introduced two improvements to the MWER loss criterion targeted at improving recognition of proper nouns.  We show that these methods improve performance of a two-pass system on several proper noun benchmarks, with WER reduction ranging from 2\% to 7\% relative.